\documentclass[manuscript]{acmart}


\usepackage{amssymb}

\usepackage[T1]{fontenc}
\usepackage{algorithm,algpseudocode}
\usepackage{afterpage}  
\usepackage{amsmath}

\newcommand{\bignegskip}{\vspace{-\bigskipamount}}

\usepackage[color]{zed}%

\usepackage{listings}
\usepackage{xcolor}
\usepackage{textcomp}
\usepackage{graphicx} 
\usepackage{comment} 
\usepackage{extpfeil} 
\usepackage{hyperref} 
\usepackage{isabelle-listings}

\definecolor{codeblue}{rgb}{0.1,0.1,0.7}
\definecolor{codegray}{rgb}{0.5,0.5,0.5}
\definecolor{codepurple}{rgb}{0.58,0,0.82}
\definecolor{backcolour}{rgb}{0.95,0.95,0.92}

\lstdefinestyle{pythonstyle}{
    backgroundcolor=\color{backcolour},
    commentstyle=\color{codegray},
    keywordstyle=\color{codeblue}\bfseries,
    stringstyle=\color{codepurple},
    basicstyle=\ttfamily\footnotesize,
    breakatwhitespace=false,
    breaklines=true,
    captionpos=b,
    keepspaces=true,
    numbers=left,
    numbersep=5pt,
    showspaces=false,
    showstringspaces=false,
    showtabs=false,
    tabsize=4
}

\lstdefinelanguage{Isabelle2}{
  morekeywords={theory, imports, begin, lemma, proof, show, have, qed, by, simp, text},
  sensitive=true,
  morecomment=[l]{--},        
  morecomment=[s]{(*}{*)},    
  morestring=[b]",
}

\lstdefinestyle{isabellestyle}{
    backgroundcolor=\color{backcolour},
    basicstyle=\ttfamily\footnotesize,
    breakatwhitespace=false,
    breaklines=true,
    captionpos=b,
    keepspaces=true,
    numbers=left,
    numbersep=5pt,
    showspaces=false,
    showstringspaces=false,
    showtabs=false,
    tabsize=2
  }

\DeclareUnicodeCharacter{21D2}{$\Rightarrow$}
\DeclareUnicodeCharacter{27F9}{$\Longrightarrow$}
\DeclareUnicodeCharacter{27F6}{$\longrightarrow$}
\DeclareUnicodeCharacter{2227}{$\wedge$}
\DeclareUnicodeCharacter{2228}{$\vee$}
\DeclareUnicodeCharacter{2200}{$\forall$}
\DeclareUnicodeCharacter{2203}{$\exists$}
\DeclareUnicodeCharacter{27E6}{$\llbracket$}
\DeclareUnicodeCharacter{27E7}{$\rrbracket$}
\DeclareUnicodeCharacter{22C0}{$\bigwedge$}
\DeclareUnicodeCharacter{2295}{$\oplus$}
\DeclareUnicodeCharacter{2204}{$\nexists$}
\DeclareUnicodeCharacter{2264}{$\le$}
\DeclareUnicodeCharacter{2265}{$\ge$}
\DeclareUnicodeCharacter{2260}{$\neq$}
\DeclareUnicodeCharacter{03B4}{$\delta$}
\DeclareUnicodeCharacter{03BB}{$\lambda$}
\DeclareUnicodeCharacter{03B5}{$\varepsilon$}
\DeclareUnicodeCharacter{2286}{$\subseteq$}
\DeclareUnicodeCharacter{2208}{$\in$}
\DeclareUnicodeCharacter{2209}{$\notin$}
\DeclareUnicodeCharacter{2261}{$\equiv$}
\DeclareUnicodeCharacter{27F7}{$\longleftrightarrow$}
\DeclareUnicodeCharacter{2987}{$\llparenthesis$}
\DeclareUnicodeCharacter{2988}{$\rrparenthesis$}
\DeclareUnicodeCharacter{22A5}{$\bot$}
\DeclareUnicodeCharacter{2218}{$\circ$}
\DeclareUnicodeCharacter{21C0}{$\rightharpoonup$}
\DeclareUnicodeCharacter{21F8}{$\pfun$}
\DeclareUnicodeCharacter{21A6}{$\mapsto$}
\DeclareUnicodeCharacter{2032}{${}^\prime$}
\DeclareUnicodeCharacter{2229}{$\cap$}
\DeclareUnicodeCharacter{222A}{$\cup$}
\DeclareUnicodeCharacter{21CC}{$\rightleftharpoons$}
\DeclareUnicodeCharacter{221E}{$\infty$} 
\DeclareUnicodeCharacter{03C3}{$\sigma$} 
\DeclareUnicodeCharacter{03C9}{$\omega$} 
\DeclareUnicodeCharacter{2211}{$\Sigma$} 
\DeclareUnicodeCharacter{03B7}{$\eta$} 

\renewcommand{\ttdefault}{txtt}

\title[Formal Analysis of the Sigmoid Function and Formal Proof of UAT]%
{Formal Analysis of the Sigmoid Function and Formal Proof of the Universal Approximation Theorem}

\author{Dustin Bryant}
\orcid{0000-0001-9876-3804}
\affiliation{%
  \institution{Independent}
  \city{Austin}
  \country{United States}
}
\email{brydustin@gmail.com}

\author{Jim Woodcock}
\orcid{0000-0001-7955-2702}
\affiliation{%
  \institution{Southwest University and State Key Laboratory of Intelligent Vehicle Safety Technology}
  \city{Chongqing}
  \country{China}
}
\affiliation{%
  \institution{Aarhus University}
  \city{Aarhus}
  \country{Denmark}
}
\affiliation{%
  \institution{University of York}
  \city{York}
  \country{United Kingdom}
}
\email{jim.woodcock@york.ac.uk}

\author{Simon Foster}
\orcid{0000-0002-9889-9514}
\affiliation{%
  \institution{University of York}
  \city{York}
  \country{United Kingdom}
}
\email{simon.foster@york.ac.uk}

\begin{document}

\begin{abstract}
  This paper presents a formalized analysis of the sigmoid function and a fully mechanized proof of the Universal Approximation Theorem (UAT) in Isabelle/HOL, a higher-order logic theorem prover. The sigmoid function plays a fundamental role in neural networks; yet, its formal properties, such as differentiability, higher-order derivatives, and limit behavior, have not previously been comprehensively mechanized in a proof assistant. We present a rigorous formalization of the sigmoid function, proving its monotonicity, smoothness, and higher-order derivatives. We provide a constructive proof of the UAT, demonstrating that neural networks with sigmoidal activation functions can approximate any continuous function on a compact interval. Our work identifies and addresses gaps in Isabelle/HOL's formal proof libraries and introduces simpler methods for reasoning about the limits of real functions. By exploiting theorem proving for AI verification, our work enhances trust in neural networks and contributes to the broader goal of verified and trustworthy machine learning.
\end{abstract}




\maketitle


\section{Introduction}
\label{sec:introduction}

Machine learning now underpins safety-critical systems, including autonomous vehicles, medical diagnostics, and robotics. In these domains, informal proofs no longer suffice. A core prerequisite for trustworthy AI is a machine-checked account of expressiveness: can our network architectures approximate the functions we need, and can we trust the proof that says so? The Universal Approximation Theorem (UAT) answers the first question, but its classical proofs are prose-based, often non-constructive (e.g. \cite{Cybenko1989}\cite{Funahashi1989}\cite{HornikStinchcombeWhite1989}) , and challenging to reuse inside verification workflows.

This paper addresses that gap. We provide, to our knowledge, the first fully mechanized, constructive proof of the UAT in a major proof assistant (Isabelle/HOL)\cite{Bryant-Woodcock-Foster-AFP2025}. Beyond validating every inference step, our constructive development yields executable witnesses, making the result directly usable within verified AI pipelines. To support this, we build missing analysis infrastructure in Isabelle to simplify limit proofs, together with a reusable formal theory of the sigmoid function, including monotonicity, smoothness, and a closed-form  derivative (via Stirling numbers). The resulting libraries are designed for reuse across mathematics, physics, engineering, and verified machine learning.

\subsection{Contributions}

\begin{enumerate}

\item \textbf{Mechanized UAT (constructive).} A complete, machine-checked proof in Isabelle/HOL that removes hidden assumptions and sets a benchmark for formalizing foundational ML results.

\item \textbf{Sigmoid calculus.} A reusable, verified development of the sigmoid's properties, including a closed-form $n$-th derivative and smoothness results.

\item \textbf{Real-analysis extensions for Isabelle.} 
Limit tools that make subsequent formalizations more natural.

\item \textbf{Path to trustworthy neuro-symbolic systems.} By embedding expressivity guarantees inside a logical framework, our results bridge symbolic and subsymbolic AI and provide a baseline for integrating expressivity verification (UAT) with robustness assurance (e.g., adversarial safety, generalization bounds).

\end{enumerate}

\subsection{Significance: Why read this paper?}

It delivers a first-of-its-kind, machine-checked, constructive UAT; strengthens the foundations of trustworthy AI; extends Isabelle/HOL with analysis tools that others can reuse; and bridges symbolic and subsymbolic AI, so that verification frameworks can rely on formally proved approximation power.

\subsection{Paper organization.}

Sect.~\ref{sec:related-work} discusses related work: relates the work to existing research by reviewing classical and constructive proofs of the Universal Approximation Theorem, prior analyses of the sigmoid function and its derivatives, and related formal-methods efforts, thereby motivating the need for a fully mechanized, reusable Isabelle/HOL formalization of these foundational results.
Sect.~\ref{sec:formalizing-the-sigmoid-function} formalizes the sigmoid function in Isabelle/HOL, defining it, proving its monotonicity, smoothness, higher-order derivatives (via Stirling numbers), and limit behavior, to establish it as a mathematically sound, sigmoidal activation function suitable for the later mechanized proof of the Universal Approximation Theorem.
Sect.~\ref{a-formal-approach-to-higher-order-derivatives-and-limit-theorems-in-isabelle-hol} extends Isabelle/HOL's analysis libraries by introducing a helpful bridge for $\varepsilon$–$N$ style formulation of limits, creating the formal infrastructure needed for reasoning about smoothness and limits in the mechanized proof of the Universal Approximation Theorem. 
Sect.~\ref{the-universal-approximation-theorem} presents the mechanized proof of the Universal Approximation Theorem in Isabelle/HOL, showing that any continuous function on a compact interval can be uniformly approximated by a finite linear combination of shifted and scaled sigmoidal functions, thereby establishing the expressive power of single-layer neural networks. That section concludes by illustrating through a worked example how the theorem applies on an actual function. 
Sect.~\ref{sec:conclusion} concludes by highlighting that the mechanization of the sigmoid function and the Universal Approximation Theorem in Isabelle/HOL eliminates informal assumptions, enhances trust in neural network theory, extends Isabelle's real analysis toolkit, and lays a rigorous foundation for future research in verified and trustworthy AI.

\section{Related Work}
\label{sec:related-work}

\subsection{Universal Approximation.}

Classical results show that shallow feed-forward networks are dense in $C(K)$ under broad conditions: Cybenko established universality for sigmoids~\cite{Cybenko1989}; Hornik generalized the picture \cite{Hornik1991}; and Leshno–Lin–Pinkus–Schocken proved that any \emph{non-polynomial} activation yields universality \cite{LeshnoLPS1993}; see also Pinkus's survey \cite{Pinkus1999}. Constructive refinements give explicit networks and bounds (e.g., \cite{ChenCL1992,CostarelliS2013}), but these proofs are typically on paper, where subtle assumptions can remain implicit. Our work provides a \emph{machine-checked, constructive} UAT in Isabelle/HOL, turning this cornerstone into reusable, verified mathematics.

\subsection{Formal Methods and Verified ML.}

Isabelle/HOL offers deep support for analysis (e.g., \cite{NipkowWP2002}), yet a convenient package for a compact UAT proof is wanting. In parallel, ML verification has advanced through DNN analyzers (Reluplex~\cite{KatzBDJK2017}, Marabou~\cite{KatzHIJLLSYWZDKB2019,WuKBK2020,WuIZTDKRAJBHLWZKKB2024}) and formal learning theory in Coq and Lean (e.g.,~\cite{BagnallS2019,TassarottiVBT2021,VajjhaTBT2021}). We extend Isabelle's toolkit with reusable formalizations and alternative limit reasoning that these lines of work can build upon.

\subsection{Activation Functions and Verified AI.}

While much verification targets piecewise-linear activations (ReLU, Leaky ReLU \cite{NairH2010,GlorotBB2011,MaasHN2013}), the sigmoid remains central in theory, probabilistic modeling\cite{Murphy-2022}, and is and remains canonical in formulations of the Universal Approximation Theorem \cite{Goodfellow-et-al-2016}. Comprehensive surveys map a vast design space (e.g., \cite{DubeySC2022,KuncK2024}{; see also \cite{Datta2020} for links to initialization), reinforcing the value of a carefully \emph{formalized} smooth activation. Our development closes this gap by verifying the higher-order properties of the sigmoid function—including a closed-form expression for its $n^{\mathrm{th}}$
 derivative via Stirling numbers—thereby enabling principled symbolic differentiation and smooth analysis in verified machine learning. Together, these results mark a step toward integrating numerical approximation theorems into the broader ecosystem of formally verified mathematics and machine learning.

\section{Formalizing the Sigmoid Function}
\label{sec:formalizing-the-sigmoid-function}

\subsection{Core Definition and Basic Properties}

The \emph{sigmoid function}, $\sigma:\mathbb{R}\to\mathbb{R}$, is defined as follows:

$$\sigma(x) = \frac{e^x}{1+e^x}\,,$$

\begin{lstlisting}[style=isabellestyle, language=Isabelle, caption=Definition of Sigmoid ] 
definition sigmoid :: "real ~⇒~ real" where 
"sigmoid = (~λ~ x::real. exp(x) / (1 + exp(x)))"
\end{lstlisting}

\noindent and a simple computation shows that an equivalent representation for the sigmoid is $\sigma(x) = \dfrac{1}{1+e^{-x}}$. 
\begin{lstlisting}[style=isabellestyle, language=Isabelle, caption=Alternate Representation of Sigmoid ] 
lemma sigmoid_alt_def: "sigmoid x = inverse (1 + exp(-x))"
proof -
  have "sigmoid x = (exp(x) * exp(-x)) / ((1 + exp(x))*exp(-x))"
    unfolding sigmoid_def by simp
  also have "... = 1 / (1*exp(-x) + exp(x)*exp(-x))"
    by (simp add: distrib_right exp_minus_inverse)
  also have "... = inverse (exp(-x) + 1)"
    by (simp add: divide_inverse_commute exp_minus)
  finally show ?thesis
    by simp
qed
\end{lstlisting}

From the definition of $\sigma(x)$ it is clear that $0< \sigma(x) <1$.   The alternate representation shows that $\sigma$ is increasing and $\sigma(0) = 1/2$. 

\begin{figure}[ht]
  \bignegskip
  \centering
  \includegraphics[width=0.25\textwidth]{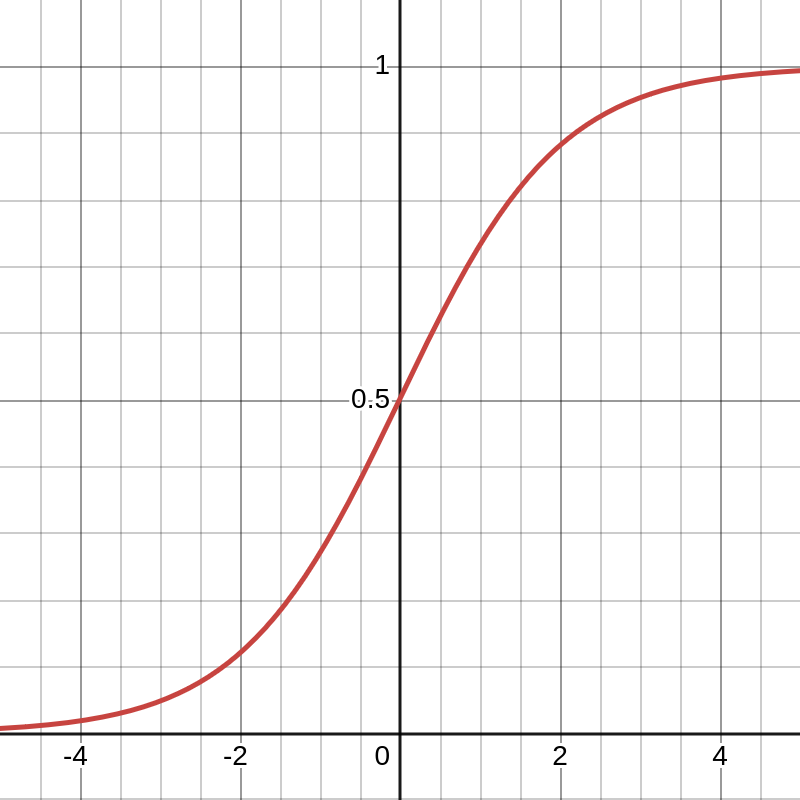}
  \caption{Sigmoid Function}
  \Description{A line plot of the logistic sigmoid function $\sigma(x)=\frac{1}{1+e^{-x}}$. The curve is S-shaped, increases monotonically from near 0 to near 1, and passes through $(0,0.5)$ with horizontal asymptotes at $y=0$ and $y=1$.}
  \label{fig:sample}
  \bignegskip
  \bignegskip
\end{figure}

\section{Formal proof of \texorpdfstring{$\sigma$}{sigma}}

Central to the learning aspect of deep learning is computing the derivative of the activation function. In the case of the sigmoid function, we have the following formulae for the first two derivatives:

$$
\begin{aligned}
    \sigma'(x) &= \sigma(x)\left[1-\sigma(x)\right]\\
    \sigma''(x) &= \sigma(x)\left[1-\sigma(x)\right]\left[1-2\sigma(x)\right]\,.
\end{aligned}
$$

\noindent Deriving these equations involves a rather straightforward calculation with the chain rule. Perhaps less well-known is the general representation for the $n^{\mathrm{th}}$ derivative of the sigmoid function(see~\cite{MinaiW1993}):

$$\sigma^{(n)}(x) \;=\; \sum_{k=1}^{n+1} (-1)^{k+1}\,(k-1)! \, S(n+1,k)\, (\sigma(x))^k\, $$

\noindent where $S(\cdot,\cdot)$ denotes a Stirling number of the second kind, formally defined in Isabelle/HOL as part of \texttt{Stirling.thy}\cite{StirlingThy-Isabelle}.  This can easily be proved by induction and amounts to splitting this into two summations, re-indexing one of the summations, and using the fact that 
$$k!S(n+1,k) + (k-1)!S(n+1,k-1) = (k-1)!S(n+2,k)\,.$$ We briefly mention the formal statement of this result:

\begin{lstlisting}[style=isabellestyle, language=Isabelle, caption=Derivatives of Sigmoid ] 
theorem nth_derivative_sigmoid:
  "~⋀~x. Nth_derivative n sigmoid x = 
    (~∑~k = 1..n+1. (-1)^(k+1) * fact (k - 1) * 
    Stirling (n+1) k * (sigmoid x)^k)"
\end{lstlisting}

We note that we have used the higher-order derivative as defined in \cite{BryantHuertayMuniveFoster2026}, which recursively defines the $k$th derivative.  The above fact shows that $\sigma$ is a smooth function as each of its derivatives is continuous.  

\subsection{Limit Behaviors and Other Results}
Central to the universal approximation theorem are the so-called sigmoidal functions,  \(f\colon \mathbb{R} \to \mathbb{R}\),  which satisfy the following property:

$$
\lim_{x \to -\infty} f(x) = 0 \quad \text{and} \quad \lim_{x \to +\infty} f(x) = 1.
$$

\begin{lstlisting}[style=isabellestyle, language=Isabelle, caption=Definition of Sigmoidal ] 
definition sigmoidal :: "(real ~⇒~ real) ~⇒~ bool" where
  "sigmoidal f ~≡~ (f ~⟶~ 1) at_top ~∧~ (f ~⟶~ 0) at_bot"
\end{lstlisting}

\noindent Indeed, it is clear that $\sigma$ is sigmoidal; in the case of mechanizing this in Isabelle, we gave a direct $\varepsilon-N$ style proof. Similarly, we showed that:

$$
\lim_{x \to -\infty} \sigma'(x) = 0 \quad \text{and} \quad \lim_{x \to +\infty} \sigma'(x) = 0.
$$

\section[Isabelle/HOL Limits and \texorpdfstring{$\varepsilon$–$N$}{ε–N} Equivalence]%
{A Formal Approach to Limits in Isabelle/HOL: Equivalence with the Classical \texorpdfstring{$\varepsilon$–$N$}{ε–N} Definition}
\label{a-formal-approach-to-higher-order-derivatives-and-limit-theorems-in-isabelle-hol}

Working with limits in Isabelle/HOL is formulated via \emph{filters}. For real-valued functions on \(\mathbb{R}\), that paradigm is correct but perhaps unfamiliar or unnecessarily abstract. In this section we present a convenient ε–\(N\) characterization for the filter \(\texttt{at\_top}\) on \(\mathbb{R}\), so that proofs can be conducted in the familiar classical style while remaining definitionally equivalent to the library’s filter semantics. Consider the following real example:

\medskip
\begin{lstlisting}[style=isabellestyle, language=Isabelle, caption={Filter-based goal (unfolded)}]
lemma lim_sigmoid_infinity: "(sigmoid ~⟶~ 1) at_top"
  unfolding tendsto_def
------------------------------------------------------------------------------------------------------
Output:
proof (prove)
goal (1 subgoal):
 1. ~∀~S. open S ~⟹~ 1 ~∈~ S ~⟹~ (~∀~_F x in at_top. sigmoid x ~∈~ S)
\end{lstlisting}

\noindent \texttt{tendsto} corresponds to the arrow denoting the limit while \texttt{at\_top} specifies that the limit is as $x$ goes to  \textit{infinity}.  Notice that it is awkward to work with the proof goal produced in the output; and this requires specialized knowledge beyond the usual definitions of limits. Instead, we want something natural to capture the feel of a classical $\varepsilon - N$ style proof from real analysis. We sought a more natural framework that captures the feel of a classical $\varepsilon - N$ proof from real analysis goals, like in the following:
\begin{lstlisting}[style=isabellestyle, language=Isabelle, caption={ε–$N$ goal obtained by rewriting to the classical form}]
lemma lim_sigmoid_infinity: "(sigmoid ~⟶~ 1) at_top"
proof (subst tendsto_at_top_epsilon_def, clarify)
------------------------------------------------------------------------------------------------------
Output:
proof (state)
goal (1 subgoal):
 1. ~⋀~ ~ε~. 0 < ~ε~ ~⟹~ ~∃~N. ~∀~ x ~≥~ N. ~¦~sigmoid x - 1~¦~ < ~ε~
\end{lstlisting}

\noindent The intended classical statement for real \(f\) is:
$$
\bigl(f \xrightarrow[x \to +\infty]{} L\bigr)
\;\Longleftrightarrow\;
\forall \varepsilon>0\;\exists N\in\mathbb{R}\;\forall x\ge N:\; |f(x)-L|<\varepsilon.
$$

Therefore, we developed a set of lemmas which allowed us to compute limits such as this and others. The limit lemma corresponding to the previously mentioned limit above is \texttt{tendsto\_at\_top\_epsilon\_def} included formally below.

\begin{lstlisting}[style=isabellestyle, language=Isabelle, caption={ε–$N$ characterization at $+\infty$ on $\mathbb{R}$}]
lemma tendsto_at_top_epsilon_def:
  "(f ~⟶~ L) at_top =
   (~∀~ ~ε~ > 0. ~∃~N::real. ~∀~ x ~≥~ N. ~¦~(f x::real) - L~¦~ < ~ε~)"
  by (simp add: Zfun_def tendsto_Zfun_iff eventually_at_top_linorder)
\end{lstlisting}

\noindent With this in hand, standard $\varepsilon$–$N$ estimates discharge goals directly.
For the logistic sigmoid $\sigma(x) = \dfrac{1}{1 + e^{-x}}$, we have
$$
0 < 1 - \sigma(x) = \frac{1}{1 + e^{x}} < e^{-x}.
$$
Choosing $N = \ln(1/\varepsilon)$ then ensures that
$$
|\sigma(x) - 1| < \varepsilon \quad \text{for all } x \ge N.
$$
Thus, \((\texttt{sigmoid} \xrightarrow{} 1)\ \texttt{at\_top}\)
follows immediately in the $\varepsilon$–$N$ formulation,
while remaining definitionally equivalent to the original filter statement.  Thus, one can naturally reason about limits without being familiar with the filters paradigm, \texttt{Zfun\_def}, or any related theories; this makes research in formal proofs of real-valued functions more accessible.

In summary, our formal development shows that the sigmoid function $\sigma$ is both smooth and sigmoidal. These properties qualify it as an effective activation function, setting the stage for our formal treatment of the Universal Approximation Theorem.

\section{The Universal Approximation Theorem}
\label{the-universal-approximation-theorem}

The proof of the UAT is central to using sigmoidal functions as approximators. Sigmoidal functions behave similarly to the Heaviside function. 
To motivate this idea, we briefly consider $\sigma$ as an approximator. We can parametrize it by introducing a \textit{weight} factor,$w$, which controls how rapidly the function transitions from $\approx 0$ to $\approx 1$, and we can parametrize its center by $x_k$. More concretely, if we parametrize the sigmoid as
$$
\sigma_{w,x_k}(x) = \frac{1}{1+e^{-w(x-x_k)}} ,
$$
where \(w > 0\) is the weight parameter and \(x_k\) is a translation parameter, increasing \(w\) stretches the function so that its transition from 0 to 1 becomes sharper. This controlled steepness is key to constructing approximations by combining appropriately weighted and shifted sigmoidal functions. Now we make precise in what sense sigmoidal functions approximate Heaviside functions:

\begin{lemma}{Sigmoidal Uniform Approximation [Costarelli and Spigler]}
 
Let \( x_0, x_1, \dots, x_N \in \mathbb{R} \), \( N \in \mathbb{N}^+ \), be fixed. For every \( \varepsilon, h > 0 \), there exists \( \overline{w} := \overline{w}(\varepsilon, h) > 0 \) such that for every \( w \geq \overline{w} \) and \( k = 0,1, \dots, N \), we have
\begin{enumerate}
    \item \( \left| \sigma(w(x - x_k)) - 1 \right| < \varepsilon \), for every \( x \in \mathbb{R} \) such that \( x - x_k \geq h \);
    \item \( \left| \sigma(w(x - x_k)) \right| < \varepsilon \), for every \( x \in \mathbb{R} \) such that \( x - x_k \leq -h \).
\end{enumerate}
\end{lemma}

\begin{lstlisting}[style=isabellestyle, language=Isabelle, caption=Sigmoidal Uniform Approximation]
lemma sigmoidal_uniform_approximation:  
  assumes "sigmoidal ~σ~"
  assumes "~ε~ > 0" and "h > 0"    
  shows "~∃~ ~ω~ >0. ~∀~ w ~≥~ ~ω~. ~∀~ k<length xs.
    (~∀~x. x - xs~!~k ~≥~ h  ~⟶~ ~¦~ ~σ~ (w * (x - xs~!~k)) - 1~¦~ < ~ε~) ~∧~
    (~∀~x. x - xs~!~k ~≤~ -h ~⟶~ ~¦~ ~σ~ (w * (x - xs~!~k))~¦~ < ~ε~)"
\end{lstlisting}

\noindent The carefully constructed lemmas from the previous section made this proof nearly trivial. We are almost in a position to state the UAT, but first, we must define some of the last few definitions used in Isabelle. Recall that a function, $f:\mathbb{R}\to \mathbb{R}$, is \textit{bounded} provided there exists $M>0$ such that $f(\mathbb{R})\subset [-M,M]$. We formally defined this in Isabelle as:

\begin{lstlisting}[style=isabellestyle, language=Isabelle, caption=Bounded Function Definition]
definition bounded_function :: "(real ~⇒~ real) ~⇒~ bool" where
  "bounded_function f ~⟷~ bdd_above (range (~λ~x. ~¦~f x~¦~))"
\end{lstlisting}

\noindent Next, given any interval $[a,b]$, we need a way to partition the interval easily widened a little on the left side into $N+1$ even subintervals each of length $\dfrac{b-a}{N}$, that is we need to partition $[a-\dfrac{b-a}{N},b]$ uniformly. We defined it in Isabelle as follows:

\begin{lstlisting}[style=isabellestyle, language=Isabelle, caption=Uniform Partition Function with Example]
definition unif_part :: "real ~⇒~ real ~⇒~ nat ~⇒~ real list" where
  "unif_part a b N =
     map (~λ~k. a + (real k -1 ) * ((b - a) / real N )) [0..<N+2]"

value "unif_part (0::real) 1 4"
(* Output: [-.25, 0,  0.25, 0.5, 0.75, 1] :: real list *)
\end{lstlisting}

\noindent This definition may seem somewhat arbitrary, but it is precisely what is required to approximate a continuous function $f$ over $[a,b]$ using sigmoidal functions, which is the UAT. Finally, we are in a position to state our main result.

\begin{theorem}[Uniform Approximation by Sigmoidal Functions~\cite{CostarelliS2013}]
Let $\sigma:\mathbb{R}\to\mathbb{R}$ be a bounded, sigmoidal function, and let $f$ be a continuous function on the interval $[a,b]$ with $a<b$. Then for every $\varepsilon>0$, there exists a positive integer $N$ and a real $w>0$ such that:
$$
\left|
\sum_{k=2}^{N+1} \bigl[f(x_k) - f(x_{k-1})\bigr]\,\sigma\!\bigl(w\,(x - x_k)\bigr)
\;+\; f(a)\,\sigma\!\bigl(w\,(x - x_0)\bigr)
\;-\; f(x)
\right| \;<\;\varepsilon
$$
for all $x\in[a,b]$,
where $\{x_0,x_1,\dots,x_{N+1}\}$ is a uniform partition of the interval 
$\left[a - \tfrac{b-a}{N},\, b\right]$.
That is, $f$ can be approximated uniformly on $[a,b]$ by a finite linear combination of translates and scalings of the sigmoidal function $\sigma$.  
\end{theorem}

\textit{Remark: } This theorem shows that $f$ can be \textit{learned} by a neural network consisting of a single layer with $N+1$ neurons that use a sigmoidal activation function $\sigma$.

\begin{lstlisting}[style=isabellestyle, language=Isabelle, caption=Uniform Approximation Theorem]
theorem sigmoidal_approximation_theorem:
  assumes sigmoidal_function: "sigmoidal ~σ~"                           
  assumes bounded_sigmoidal: "bounded_function ~σ~"                       
  assumes a_lt_b: "a < b"                                               
  assumes contin_f: "continuous_on {a..b} f"                            
  assumes eps_pos: "0 < ~ε~"
  defines "xs N ~≡~ unif_part a b N"
  shows "~∃~N::nat. ~∃~(w::real) > 0.(N > 0) ~∧~
    (~∀~x ~∈~ {a..b}. ~¦~(~∑~k~∈~{2..N+1}. 
    (f(xs N~!~k) - f(xs N~!~(k - 1))) * ~σ~(w * (x - xsN~!~k)))
    + f(a) * ~σ~(w * (x - xs N~!~0)) - f x~¦~ < ~ε~)"
\end{lstlisting}

The approximant 

$$
G_{N,f}(x)
= f(a)\,\sigma\!\bigl(w(x - x_0)\bigr) + \sum_{k=2}^{N+1} \bigl(f(x_k) - f(x_{k-1})\bigr)\,\sigma\!\bigl(w(x - x_k)\bigr).
$$
 is a one-hidden-layer network where all units share the same slope \( w \)
and only their shifts \( x_k \) differ. The output weights are just forward
differences of \( f \) on a uniform grid. Two error terms arise in the proof, $I_1$ and $I_2$; $I_1$ measures how well each sigmoid acts like a step function away from its center (controlled by making $w$ large) and $I_2$ measures how much the piecewise-constant “finite-difference” reconstruction deviates from 
$f$ locally (controlled by making the mesh $h$ small and hence $N$ large).  More precisely, the proof introduces a local surrogate of $G_{N,f}$, namely $L_i$, that pretends distant sigmoids are already saturated (exactly $1$ on the left, $0$ on the right), but keeps the two boundary sigmoids at $x_i$ and $x_{i+1}$ in their nonsaturated true form.  Concretely, in our mechanization we fix $\varepsilon >0$ then we define $\eta \;=\;
\frac{\varepsilon}{
  \bigl(\text{Sup}_{x \in [a,b]} |f(x)|\bigr)
  + 2\,\bigl(\text{Sup}_{x \in \mathbb{R}} |\sigma(x)|\bigr)
  + 2
}
$, and we obtain $\delta$ so that whenever $|x-y|<\delta$ we have $|f(x)-f(y)|<\eta$. Finally, we require at least $N$ neurons with 
$N =
\left\lfloor
  \max\!\left\{
    3,\;
    \frac{2(b - a)}{\delta},\;
    \frac{1}{\eta}
  \right\}
\right\rfloor + 1.
$  With this in place we let $\{x_0,x_1, \ldots, x_{N+1}\}$ be the uniform partition of the interval $[a-\frac{b-a}{N},b]$ and note that the distance between adjacent $x_k$ is $|x_k - x_{k-1}|=\frac{b-a}{N} < \delta/2$ from our choice of $N$. Let us consider an arbitrary $x\in [a,b]$, we define $i=i(x) = \max\{i\in \{1,\ldots, N\}: x_i\le x\}$ so that $x\in [x_i,x_{i+1}]$ then for $i \ge 3$:

$$
\begin{aligned}
L_i(x)
&= \underbrace{f(a)+\sum_{k=2}^{i-1}\big(f(x_k)-f(x_{k-1})\big)}_{\text{``all steps left of }x\text{ are on''}}
\;+\;
\underbrace{(f(x_i)-f(x_{i-1}))\,\sigma\!\big(w(x-x_i)\big)}_{\text{left boundary still transitioning}}
\\[4pt]
&\quad+\;
\underbrace{(f(x_{i+1})-f(x_i))\,\sigma\!\big(w(x-x_{i+1})\big)}_{\text{right boundary still transitioning}}\,.
\end{aligned}
$$

Now we measure the error of our approximant against the objective function $f$ and split this into two terms:

$$
|G_{N,f}(x) - f(x)| \leq 
\underbrace{|G_{N,f}(x) - L_i(x)|}_{I_1} +
\underbrace{|L_i(x) - f(x)|}_{I_2}.
$$

We will make it explicit that \(I_1\) represents the cumulative
error from all the ``far'' sigmoids---that is, those which should already be saturated to
\(0\) or \(1\). After a careful  rearrangement of \(G_{N,f}(x) - L_i(x)\),
we see that:

$$
\begin{aligned}
I_1(i,x)
&= \Bigl|\,
    f(a)\bigl(\sigma(w(x - x_0)) - 1\bigr)
    + \sum_{k=2}^{i-1} \bigl(f(x_k) - f(x_{k-1})\bigr)\bigl(\sigma(w(x - x_k)) - 1\bigr) \\
&\qquad\quad
    + \sum_{k=i+1}^{N+1} \bigl(f(x_k) - f(x_{k-1})\bigr)\,\sigma\!\bigl(w(x - x_k)\bigr)
   \Bigr| \\[4pt]
&\overset{\triangle}{\le}
   |f(a)|\,\bigl|\sigma\!\bigl(w(x-x_0)\bigr)-1\bigr|
   + \sum_{k=2}^{i-1} \bigl|f(x_k)-f(x_{k-1})\bigr|\,\bigl|\sigma\!\bigl(w(x-x_k)\bigr)-1\bigr| \\
&\qquad\quad
   + \sum_{k=i+1}^{N+1} \bigl|f(x_k)-f(x_{k-1})\bigr|\,\bigl|\sigma\!\bigl(w(x-x_k)\bigr)\bigr|.\quad \dagger
\end{aligned}
$$

The next step is to bound each term using the uniform-continuity and sigmoidal
estimates. Specifically, we aim to establish
$$
I_1(i,x)
< |f(a)|\,|\sigma(w(x-x_0))-1|
  + \sum_{k=2}^{i-1} \eta\,\tfrac{1}{N}
  + \sum_{k=i+2}^{N+1} \eta\,\tfrac{1}{N}.
$$

From here, the formalization compels us to analyze a number of side cases.  First, we consider the case when $i\ge 3$.  If $\sigma(w(x-x_k)) = 0$ for every $k\in \{2,\ldots,i-1\}$ then $\dagger$ above greatly simplifies.  We further consider the case by when $i=N$.  If $i=N$ then the result follows immediately; otherwise, we next consider when the second summation's summands of $\dagger$ are all identically $0$.  In the edge case when all the summands are zero, we get the strict inequality for free; otherwise, we collect the terms where the summand is nonzero and use the fact that $|f(x_k)-f(x_{k-1})|< \eta$ by construction of the mesh of points.  Moreover, when $k<i$ we have that $|\sigma(w(x-x_k))-1|<\frac{1}{N}$ but when $k>i$ we get that $|\sigma(w(x-x_k))|<\frac{1}{N}$.  Thus, we can see that when we approximate the neural layer $G$ with terms that are mostly in their limit, we obtain a very good estimate.  We further note that checking all of these edge cases is not merely an exercise in logic; with $ \sigma(x)$ equal to a Heaviside function, it is easy to see that one of the summands could be entirely composed of zeros. Observe that we required $N>3$ as this made it possible for there to exist an $i$ such that $\{2,\ldots, i-1\}\neq \emptyset$ and $\{i+2,\ldots, N+1\} \neq \emptyset$ whenever $i\ge3$ in the main case so that the summations are nonempty.

When $i\le 2$, the argument essentially repeats itself, but we use a different definition for $L_i$.   Finally, since $\{2,\ldots, i-1,i+2,\ldots , N+1\}\subset \{2,\ldots, N+1\}$ we get that the last inequality simplifies to

$$
\begin{aligned}
I_1(i,x)
  &< |f(a)|\,\bigl|\sigma\!\bigl(w(x-x_0)\bigr) - 1\bigr| + \eta \\[4pt]
  &\le |f(a)|\,\frac{1}{N} + \eta \\[4pt]
  &< (1 + |f(a)|)\,\eta \\[4pt]
  &\le \bigl(1 + \sup_{t \in [a,b]} |f(t)|\bigr)\,\eta,
\end{aligned}
$$

Next, we need to handle the bounding of $I_2$:

$$
\begin{aligned}
I_2(i,x)
  &= \Bigl|\,
      \sum_{k=2}^{i-1} \bigl(f(x_k) - f(x_{k-1})\bigr)
      + f(a)
      + \bigl(f(x_i) - f(x_{i-1})\bigr)\,\sigma\!\bigl(w(x - x_i)\bigr) \\[2pt]
  &\qquad\quad
      + \bigl(f(x_{i+1}) - f(x_i)\bigr)\,\sigma\!\bigl(w(x - x_{i+1})\bigr)
      - f(x)
    \Bigr|.\\
    &= \Bigl|\,
      f(x_{i-1}) - f(x)
      + \bigl(f(x_i) - f(x_{i-1})\bigr)\,\sigma\!\bigl(w(x - x_i)\bigr) 
      + \bigl(f(x_{i+1}) - f(x_i)\bigr)\,\sigma\!\bigl(w(x - x_{i+1})\bigr)
    \Bigr|.\\
    &\le 
     |f(x_{i-1}) - f(x)|
     + \bigl|f(x_i) - f(x_{i-1})\bigr|\,\bigl|\sigma\!\bigl(w(x - x_i)\bigr)\bigr| 
     + \bigl|f(x_{i+1}) - f(x_i)\bigr|\,\bigl|\sigma\!\bigl(w(x - x_{i+1})\bigr)\bigr|.\\
     &< \bigl(1 + 
        \,|\sigma\!\bigl(w(x - x_i)\bigr)|
        + |\sigma\!\bigl(w(x - x_{i+1})\bigr)|      
      \bigr)\,\eta.\\
      &\le \bigl(
        2\,\text{Sup}_{x \in \mathbb{R}}\,|\sigma(x)|
        + 1
      \bigr)\,\eta.
\end{aligned}
$$

Thus, our total error is given by

$$
\begin{aligned}
|G_{N,f}(x) - f(x)|
&\le \bigl(1 + \sup_{t \in [a,b]} |f(t)|\bigr)\,\eta
   + \bigl(2\,\text{Sup}_{x \in \mathbb{R}}|\sigma(x)| + 1\bigr)\,\eta \\[4pt]
&= \Bigl(\sup_{t \in [a,b]} |f(t)|
          + 2\,\text{Sup}_{x \in \mathbb{R}}|\sigma(x)|
          + 2\Bigr)\,\eta.\\
          &< \varepsilon.
\end{aligned}
$$

The Isabelle proof of the UAT is around 1,350 lines of code and approximately two-thirds of it is devoted to proving the bounds on $I_1$ and $I_2$ in the various previously mentioned cases.  This is done, in particular, because $\texttt{sum\_strict\_mono}$ was used to establish the inequality specifically for the terms when $\sigma(w(x-x_k))\neq0$ but establishing the sets where this is the case lead to the proof being longer than anticipated.

It is often casually stated that ``the universal approximation theorem is \textit{merely} an existential claim, it does not tell you how many neurons are necessary to estimate your function''. While this is true, it is somewhat misleading.  Cybenko's classic, nonconstructive proof \cite{Cybenko1989} gave no bound on the number of neurons necessary; however, Costarelli's constructive proof shows us how many neurons are sufficient, provided certain terms can be determined.  In general, it might be possible to use fewer neurons than implied by the constructive proof, but the proof essentially gives us an recipe for a sufficient number of neurons.  Pick conservative bounds $M_f \ge \sup_{t\in[a,b]}|f(t)|$ and $M_\sigma \ge \sup_{x\in\mathbb{R}}|\sigma(x)|$, for example $M_\sigma = 1$ for sigmoid, and let $\delta$ be such that $\left|x-y\right|< \delta$ implies $\left|f(x)-f(y)\right|<\eta$ on $[a,b]$, if $f$ is L-Lipschitz then $\delta=\eta/L$ works.  Finally, pick suffiicently large $w$, if $\sigma$ is the sigmoid then $w=\frac{\ln(N-1)}{h}$ suffices.  With these choices the definition given for $N$ above will suffice for the number of neurons such that the universal approximation theorem is satisfied for $f$.

\section{Conclusion}
\label{sec:conclusion}

We addressed several fundamental questions: How can the sigmoid function and its higher-order derivatives be formally defined and analyzed? How does the UAT ensure that neural networks with sigmoidal activation functions can approximate any continuous function on a compact interval? How can these results be mechanized in Isabelle/HOL to provide a fully verified, constructive proof of function approximation? In answering these questions, we uncovered gaps in Isabelle's formal proof libraries, particularly in limit reasoning. This formality led us to introduce alternative limit formulations.

Isabelle proved particularly insightful in this endeavor, as it enforced a mathematical rigor that traditional proofs often lack. It eliminated informal reasoning, ensured the correctness of complex derivations (such as the $n^{\mathrm{th}}$ derivative of the sigmoid function using Stirling numbers), and provided a constructive proof of the UAT, making it directly applicable to verified AI. Furthermore, our work enhanced Isabelle's real analysis toolkit, improving its usability for future machine learning and mathematical analysis formalizations.

This paper demonstrates the power of theorem proving in AI verification, reinforcing the trustworthiness of neural network models. By bridging formal methods, real analysis, and deep learning, our work lays the groundwork for future research in verified machine learning and trustworthy AI systems.

\nocite{*}
\bibliographystyle{plainurl}
\bibliography{bibliography}

\section*{Appendix I: A Worked Example Bounding the Neurons Needed}
Now, let us consider the following real example.  On $[0,1]$, let
$$
f(x) \;=\; \left|x-0.3\right| \;+\; 0.3\sin\left(6\pi x\right) \;+\; 0.2\,x\left(1-x\right)
$$
and we wish to model this with the sigmoid function. 
We have $M_f\le 0.7+0.3+0.05=1.05$ and $M_\sigma=1$.
For $\varepsilon=10^{-2}$,$$
\eta=\frac{0.01}{(1+1.05)+(2\cdot 1+1)}=\frac{0.01}{5.05}\approx 1.98\times 10^{-3}.
$$
A Lipschitz bound is $L:=1 + 0.3\cdot 6\pi + 0.2 \approx 6.855$, hence $
\delta=\eta/L \approx 2.89\times 10^{-4}.$
Choose 
$$
N=\Bigl\lfloor \max\{\,3,\;2/\delta,\;1/\eta\,\}\Bigr\rfloor + 1
 = \Bigl\lfloor \max\{3,\; 2/(2.89\times 10^{-4}),\; 505.05\}\Bigr\rfloor + 1
 = 6,925
$$
so $h=1/N\approx 1.444\times 10^{-4}$ and $1/N\le \eta$.
Finally,
$$
w \;\ge\; \frac{\ln(N-1)}{h}
     \;=\; \frac{\ln 6924}{1/6925}
     \;\approx\; 61,237.
$$
With these choices, $\sup_{x\in[0,1]}|G_{N,f}(x)-f(x)|< 10^{-2}$.

\section*{Appendix II: Proving the recipe bound for $N$}

We use the sigmoid
$
\sigma(x) = \frac{1}{1+e^{-x}}.
$
and want
$
\lvert \sigma(wt) - 1 \rvert \le \frac{1}{N}
\quad\text{for all } t \ge h.
$
Since $\sigma(wt) \le 1$, this is equivalent to
$
1 - \sigma(wt) \le \frac{1}{N}.
$

For any $t \ge h$,
$$
\begin{aligned}
1 - \sigma(wt)
&= 1 - \frac{1}{1+e^{-wt}} \\
&= \frac{1+e^{-wt}}{1+e^{-wt}} - \frac{1}{1+e^{-wt}} \\
&= \frac{e^{-wt}}{1+e^{-wt}} \\
&= \frac{1}{1+e^{wt}}.
\end{aligned}
$$

Thus the condition $1 - \sigma(wt) \le 1/N$ is
$$
\frac{1}{1+e^{wt}} \le \frac{1}{N}.
$$
Because \(N > 0\) and \(e^{wt} > 0\), this is equivalent to
$$
\begin{aligned}
N &\le 1 + e^{wt}, \\
N - 1 &\le e^{wt}, \\
\ln(N - 1) &\le wt, \\
w &\ge \frac{1}{t}\ln(N - 1).
\end{aligned}
$$

So for fixed \(N\),
$$
1 - \sigma(wt) \le \frac{1}{N}
\quad\Longleftrightarrow\quad
w \ge \frac{1}{t}\ln(N - 1).
$$

To make this hold for all $t \ge h$, we require
$$
w \ge \sup_{t \ge h} \frac{1}{t}\ln(N - 1).
$$
Since $\frac{1}{t}\ln(N - 1)$ is decreasing in $t > 0$, the supremum over $t \ge h$ is attained at $t = h$, hence
$$
\sup_{t \ge h} \frac{1}{t}\ln(N - 1) = \frac{1}{h}\ln(N - 1).
$$

Therefore it suffices to choose
$$
w = \frac{1}{h}\ln(N - 1).
$$

Now we use the Lipschitz continuity of $f$ to obtain an explicit
modulus of continuity. Suppose that $f$ is $L$-Lipschitz on $[a,b]$, i.e.
$$
  |f(x) - f(y)| \le L\,|x-y|
  \qquad \text{for all } x,y \in [a,b].
$$
Given any $\eta > 0$, we seek $\delta > 0$ such that
$$
  |x-y| < \delta
  \;\Longrightarrow\;
  |f(x) - f(y)| < \eta
  \qquad \text{for all } x,y \in [a,b].
$$
Using the Lipschitz condition, whenever $|x-y| < \delta$ we have
$$
  |f(x) - f(y)|
  \le L\,|x-y|
  < L\,\delta.
$$
Thus it suffices to choose $\delta$ so that $L\,\delta \le \eta$, and one
convenient choice is
$$
  \delta = \frac{\eta}{L}.
$$
With this choice, for all $x,y \in [a,b]$ satisfying $|x-y| < \delta$ we obtain
$$
  |f(x) - f(y)|
  \le L\,|x-y|
  < L\left(\frac{\eta}{L}\right)
  = \eta,
$$
which is exactly the uniform continuity condition required in the construction.

\end{document}